# COOPERATIVE TWO-QUANTUM INTERACTION OF EXCITED SYSTEM WITH BATH

**Nicolae A. Enaki**[a] and **Vitalie Eremeev**[a,b]

[a]Institute of Applied Physics, Academy of Sciences of Moldova,
Academiei str.5, Chisinau MD-2028, Moldova
E-mail: enache@as.md

[b]Department of Engineering and Computer Sciences, Free International University of Moldova,
Vlaicu Parcalab str.52, Chisinau MD-2012, Moldova

## 1. Introduction

The problem of the two-photon coherent generation of entanglement photon pairs in quantum optics has been intensively studied for the last years. It is important to note that the two-quantum cooperative effects play a main role in other fields of physics as well. The collective processes in condense matter have many analogical proprieties with cooperative radiation effects in quantum optics[1,2]. Recently the two-quantum generation of entanglement photons and its applications in information technology and communication has been intensively studied[3,4]. The new cooperative emission phenomenon for dipole-forbidden transitions of inverted system of radiators can be observed in the processes of the two-photon spontaneous emission [5,6].

It is interesting from the physical point of view to study the cooperative phenomena in Statistical physics in more large aspects. In proposed paper it is studied the cooperative phase-transition in the system of radiators that interact with the thermostat through the two-quantum processes. Thus, this is possible when the one-photon interaction with the thermostat is forbidden. The temperature dependence of the super-radiant order parameter in this situation is investigated. As the exchange integral between the radiators strongly depends on the temperature at first stage we observe an enhancement of the order parameter as temperature function. Achieving the maximal value, the order parameter decreases till zero value in the similar way as in usual temperature dependence of the order parameter in one-photon super-radiance. The master equation and the stationary solution, which describes the two-photon exchange of radiators with thermal bath, are obtained. The possible diagram of the two-photon exchange between the radiators and thermostat is briefly analyzed.

One other example is superconductivity, where the Cooper-pairs are created due to the simultaneous two-phonon exchange between electrons. It occurs when the one-phonon exchange integral between the band electrons is smaller than the two-phonon exchange. This is possible in many-band superconducting materials, in which the two-phonon exchange integral arises through the virtual bands of material. Some estimates of the two-phonon super-conductivity have already been proposed. In many cases this effect was studied decomposing the interaction Hamiltonian on constant of the electron-phonon interaction[7]. A more realistic model that takes into account the specificities of the many-band aspects of superconductor materials will be proposed in this paper. In two-phonon processes, a more complicated temperature dependence of the order parameter is expected. In a rigorous study of this anomalous temperature dependence of the order parameter of superconductors is presented. In the proposed model one expects that the two-phonon exchange effects may amplify the superconductivity in a way similar to that the thermal field amplifies the two-photon super-radiance in a micro-cavity. In this paper it is studied the two-photon super-radiance in micro-cavity and two-phonon superconductivity in many-band materials.

## 2. Two-photon super-radiance

Let us to consider an ensemble of radiators in the micro-cavity that enters into two-photon resonance with cavity electromagnetic field. Using the adiabatical method of elimination of EMF operators one can obtain the following master-equation for atomic system[6].

$$\frac{d\rho}{dt} = i(\omega_{21} + 2n^2\chi)[\rho, R_z] + i(1+2n)\chi[\rho, R^+R^-] - (1+n)^2\gamma(\rho R^+R^- - 2R^-\rho R^+ + R^+R^-\rho)$$
$$- n^2\gamma(\rho R^-R^+ - 2R^+\rho R^- + R^-R^+\rho). \qquad (2.1)$$

Here we have the following parameters: $\chi = \frac{A(k)}{\hbar^4} \frac{\omega_{21} - 2\omega_k}{(2\omega_k - \omega_{21})^2 + 4\Gamma^2}$,

$\gamma = \frac{A(k)}{\hbar^4} \frac{2\Gamma}{(2\omega_k - \omega_{21})^2 + 4\Gamma^2}$. The parameter $\chi$ describes the cooperative atom-atom interaction through cavity EMF vacuum; $\gamma$ is a spontaneous emission rate of single atom in the cavity. This parameter is expressed through the detuning $\delta = \omega_{21} - 2\omega_k$ and damping factor of micro-cavity $\Gamma$. In Eq.(2.1) is introduced the following quasi-spin operators $R^\pm = \sum_{j=1}^N R_j^\pm$, $R_z = \sum_{j=1}^N R_{zj}$ obeying the commutation relations: $[R^+, R^-] = 2R_z$, $[R_z, R^\pm] = \pm R^\pm$.

It is observed that for large values of $\delta = \omega_{21} - 2\omega_k$ the value of $\gamma$ decreases, while the absolute value of $\chi$ increases. In this case we observe that the part of Hamiltonian, which describes the two-photon atom-atom interaction, increases. This effective Hamiltonian has the following form

$$H^{eff} = \hbar\omega_0 R_z + \hbar\chi(1+2n)R^+R^-, \qquad (2.2)$$

where $\omega_0 = \omega_{21} + 2n^2\chi$ and $\hbar\chi(1+2n)$ is the energetic exchange integral between the radiators. In comparison with one-photon super-radiance phase transition in this case the exchange integral depends on temperature through the mean number of photons $n = 1/[\exp(\hbar\omega_k/\kappa_B T) - 1]$, where $\kappa_B$ is Boltzmann constant.

Using the conservation of Bloch-vector at low temperature $R^2 = R_z^2 - R_z + R^+R^-$, (where $R$ is a constant), we can express Hamiltonian (2.2) through the operator $R_z$

$$H^{eff} = \hbar\{\omega_0 + \chi(1+2n)\}R_z + \hbar\chi(1+2n)(R+R_z)(R-R_z) \qquad (2.3)$$

From the condition of the minimum value of the Hamiltonian it is obtained the mean value of $R_z$

$$\langle R_z \rangle = \frac{1}{2} + \frac{\omega_0}{2\chi(1+2n)} \qquad (2.4)$$

From this relation and condition of conservation of Bloch-vector follows that $\langle R_z \rangle$ must take the negative values: $-j < \langle R_z \rangle < 0$. Here $j=N/2$ the maximal value of Bloch-vector, $N$ is the number of atoms in the micro-cavity. For a large number of radiators the two-photon exchange integral $\chi$ must satisfy the inequality: $-\omega_0/[2j(1+2n)] < \chi < 0$. For large value of $j$ the quasi-spin system can be approximated by the following Hamiltonian[8]

$$H^{eff} = \hbar\omega_0 R_z + \frac{1}{2}\hbar\chi(1+2n)C(R^+ + R^-), \qquad (2.5)$$

where $C = \langle R^- \rangle = \langle R^+ \rangle$.

The Hamiltonian (2.5) is obtained from (2.2) using the following approximation: $2R^+R^- = \langle R^+ \rangle R^- + \langle R^- \rangle R^+$.

Below we are interested in phase transition of quasi-spin system from correlated state to uncorrelated state stimulated by increase of temperature. In order to found the equation for order parameter C we propose the following transformations[9]

$$R^+ = \alpha S^+ + \beta S^- + \gamma S_z, \quad R_z = -\frac{\gamma}{2}(S^+ + S^-) + (\alpha + \beta)S_z \qquad (2.6)$$

where $\alpha = (1+\sqrt{v})/2$, $\beta = -(1-\sqrt{v})/2$, $\gamma = \sqrt{1-v}$, $v = \frac{\omega_0^2}{\omega_0^2 + \Omega^2}$, $\Omega = \chi(1+2n)C$,

In this case Hamiltonian (2.5) takes the diagonal form

$$H^{eff} = \hbar\{(\alpha + \beta)\omega_0 + \gamma\Omega\}S_z = \hbar\overline{\omega}S_z, \qquad (2.7)$$

where $\overline{\omega} = \sqrt{\omega_0^2 + \Omega^2}$

Using this Hamiltonian, it is obtained the following equation for the order parameter

$$C \equiv \langle R^- \rangle = \frac{Sp\{\exp[-H^{eff}/(\kappa_B T)]R^-\}}{Sp\{\exp[-H^{eff}/(\kappa_B T)]\}}, \qquad (2.8)$$

$$|C| = \frac{\Omega}{2\overline{\omega}}\{\cot\tanh(\hbar\overline{\omega}/2\kappa_B T) - (2j+1)\cot\tanh(\hbar\overline{\omega}(2j+1)/2\kappa_B T)\} \qquad (2.9)$$

From numerical and analytical results follows that such two-photon cooperative phase transition has some particularities comparatively with one-photon super-radiance phase transition (see Fig.2). One of particularity consists in dependence of exchange integral $\hbar\chi(1+2n)$ on the temperature through the mean number of photons.

## 3. Two-phonon superconductivity

In this section, the two-band electronic system, which interacts with the phononic field, is proposed for theoretical discussion. For the simplicity, the Coulombic interaction between electrons is not taken into consideration. Therefore the Hamiltonian of such a system may be represented through the two-phonon interaction between the electrons of lower band in the following form[10]

$$H = \sum_k \varepsilon_1'(\mathbf{k})a_k^+ a_k - \frac{1}{V}\left\{\sum_{k_1}\sum_{\mathbf{q},\mathbf{q}_1} g_{2,1}(\mathbf{q})g_{2,1}^*(\mathbf{q}_1)\frac{a_{1,k+\mathbf{q}-\mathbf{q}_1}^+ a_{1,k}b_{\mathbf{q}_1}^+ b_{\mathbf{q}}}{\varepsilon_2(\mathbf{k}+\mathbf{q}) - \varepsilon_1(\mathbf{k}+\mathbf{q}-\mathbf{q}_1) - \hbar\omega_{\mathbf{q}_1}}\right.$$

$$\left. + g_{2,1}(\mathbf{q})g_{2,1}(\mathbf{q}_1)\frac{a_{1,k+\mathbf{q}+\mathbf{q}_1}^+ a_{1,k}b_{\mathbf{q}_1}b_{\mathbf{q}}}{\varepsilon_2(\mathbf{k}+\mathbf{q}) - \varepsilon_1(\mathbf{k}+\mathbf{q}+\mathbf{q}_1) + \hbar\omega_{\mathbf{q}_1}} + h.c\right\} \qquad (3.1)$$

Here $V$-is the volume of the crystal; $b_{\mathbf{q}}^+(b_{\mathbf{q}})$ is the boson creation (annihilation) operator with $\mathbf{q}$ wave-vector, and $a_{n,k}^+(a_{n,k})$ is the electron creation (annihilation) operator in $m$-band ($m=1,2$) with the quasi-wave vector - $\mathbf{k}$ and spin - $\sigma$. Therefore the sums on $k$ in Hamiltonian (3.1) take into account the wave vector and the spin of electrons in the bands, $k=(\mathbf{k},\sigma)$; $\varepsilon_m(\mathbf{k})$ is the energy of the electron, which can be calculated from the level of the chemical potential. The Hamiltonian (3.1) was obtained with the consideration that interband matrix element of the electron-phonon interaction $g_{12}$ is larger than intraband matrix elements $g_{11}$, $g_{22}$, i.e. $g_{12} \gg g_{11}$ ($g_{22}$). The matrix elements of the electron-phonon interaction are known[11].

After the elimination of bi-boson operators it is obtained the effective Hamiltonian that describes the two-quanta interaction between the electrons

$$H^{eff} = -\frac{1}{V^2} \sum_{k,k_1} \sum_{\mathbf{q},\mathbf{q_1}} |g_{21}(\mathbf{q})|^2 |g_{21}(\mathbf{q_1})|^2 (1+2N_{\mathbf{q_1}}) \phi_1(\mathbf{k},\mathbf{q},\mathbf{q_1}) \phi_1(\mathbf{k_1},\mathbf{q},\mathbf{q_1}) \frac{(\hbar\omega_{\mathbf{q}} + \hbar\omega_{\mathbf{q_1}}) a^+_{k+\mathbf{q}+\mathbf{q_1}} a^+_{k_1} a_{k_1+\mathbf{q}+\mathbf{q_1}} a_k}{(\hbar\omega_{\mathbf{q}} + \hbar\omega_{\mathbf{q_1}})^2 - (\varepsilon_1(\mathbf{k_1}+\mathbf{q}+\mathbf{q_1}) - \varepsilon_1(\mathbf{k_1}))^2}$$

$$-\frac{2}{V^2} \sum_{k,k_1} \sum_{\mathbf{q},\mathbf{q_1}} |g_{21}(\mathbf{q})|^2 |g_{21}(\mathbf{q_1})|^2 N_{\mathbf{q_1}} \phi_2(\mathbf{k},\mathbf{q},\mathbf{q_1}) \phi_2(\mathbf{k_1},\mathbf{q},\mathbf{q_1}) \frac{(\hbar\omega_{\mathbf{q}} - \hbar\omega_{\mathbf{q_1}}) a^+_{k+\mathbf{q}-\mathbf{q_1}} a^+_{k_1} a_{k_1+\mathbf{q}-\mathbf{q_1}} a_k}{(\hbar\omega_{\mathbf{q}} - \hbar\omega_{\mathbf{q_1}})^2 - (\varepsilon_1(\mathbf{k_1}+\mathbf{q}-\mathbf{q_1}) - \varepsilon_1(\mathbf{k_1}))^2} ; \quad (3.2)$$

where $\phi_1(\mathbf{k_1},\mathbf{q},\mathbf{q_1}) = \frac{\varepsilon_2(\mathbf{k}+\mathbf{q}) + \varepsilon_2(\mathbf{k}+\mathbf{q_1}) - \varepsilon_1(\mathbf{k}+\mathbf{q}+\mathbf{q_1}) - \varepsilon_1(\mathbf{k})}{(\varepsilon_2(\mathbf{k}+\mathbf{q}) - \varepsilon_1(\mathbf{k}+\mathbf{q}+\mathbf{q_1}) + \hbar\omega_{\mathbf{q_1}})(\varepsilon_2(\mathbf{k}+\mathbf{q_1}) - \varepsilon_1(\mathbf{k}) - \hbar\omega_{\mathbf{q_1}})}$,

$\phi_2(\mathbf{k_1},\mathbf{q},\mathbf{q_1}) = \frac{\varepsilon_2(\mathbf{k}+\mathbf{q}) + \varepsilon_2(\mathbf{k}-\mathbf{q_1}) - \varepsilon_1(\mathbf{k}+\mathbf{q}-\mathbf{q_1}) - \varepsilon_1(\mathbf{k})}{(\varepsilon_2(\mathbf{k}+\mathbf{q}) - \varepsilon_1(\mathbf{k}+\mathbf{q}-\mathbf{q_1}) - \hbar\omega_{\mathbf{q_1}})(\varepsilon_2(\mathbf{k}-\mathbf{q_1}) - \varepsilon_1(\mathbf{k}) + \hbar\omega_{\mathbf{q_1}})}$.

As is observed from the effective interaction Hamiltonian (3.2) the transitions between the two states of the first band can take place with absorption of one phonon with $\mathbf{q}$ wave-vector and emission of the other phonon with $\mathbf{q_1}$ wave-vector. Thus, the two-phonon cooperative exchange between the electrons increases with increasing of the temperature. This is one of the main differences between the one-phonon Bardeen-Cooper-Schrieffer exchange and two-phonon exchange between the electrons. Such temperature dependence is very important for the formation of the superconducting phase. The influence of the second virtual band position on the creation of the two-phonon Cooper effect in the first band plays an important role in this model.

In this model it is considered that $\phi_1(\mathbf{k},\mathbf{q},\mathbf{q_1}) \approx \phi_2(\mathbf{k},\mathbf{q},\mathbf{q_1}) \approx 2/\varepsilon_{21}$, where $\varepsilon_{21}$ is the energetic distance from the Fermi surface to the bottom of the second band. After a simple modification of the effective Hamiltonian (3.2) the following interaction Hamiltonian between the electrons is obtained

$$H = \sum_k \varepsilon'_1(\mathbf{k}) a^+_k a_k - \frac{G(T)}{V} \sum_{k,k_1} a^+_{k_1} a^+_{-k_1} a_{-k} a_k , \quad (3.3)$$

where: $\pm \mathbf{k} = \pm \left(\mathbf{k}, \frac{1}{2}\right)$ and $G(T) = \frac{4}{V} \sum_{\mathbf{q_1}} \frac{|g_{21}(\mathbf{q}-\mathbf{q_1})|^2 |g_{21}(\mathbf{q_1})|^2}{\varepsilon_{21}^2} \left\{ \frac{1}{\hbar\omega_{\mathbf{q}-\mathbf{q_1}}} + 4N_{\mathbf{q_1}} \frac{\hbar\omega_{\mathbf{q}-\mathbf{q_1}}}{(\hbar\omega_{\mathbf{q}-\mathbf{q_1}})^2 - (\hbar\omega_{\mathbf{q_1}})^2} \right\} \quad (3.4)$

Introducing the superconductivity order parameter $\Delta(T) = \frac{G(T)}{2V} \sum_{k_1} \langle a^+_{k_1} a^+_{-k_1} \rangle$, we can approximate the Hamiltonian (3.3) in the following form:

$$H = \sum_k \varepsilon'_1(\mathbf{k}) a^+_k a_k - \sum_k (\Delta(T) a_{-k} a_k + h.c.) \quad (3.5)$$

After the intermediary transformations in (3.4) one can obtain the final expression for $G(T)$

$$G(T) = G^{(0)}(1+\gamma); \quad G^{(0)} = \frac{|M_{21}|^4}{\pi^2 \varepsilon_{21}^2} \frac{1}{\hbar c} \ln\left|\frac{q_{max}}{k_F}\right|, \quad \gamma = \left(\alpha \ln\left|\frac{q_{max}}{k_F}\right|\right)^{-1} \sum_{n=1}^{\infty} \frac{1}{n} \{\exp(-n\alpha) Ei(n\alpha) - \exp(n\alpha) Ei(-n\alpha)\} \quad (3.6)$$

Here $\alpha = \hbar c k_F /(\kappa_B T)$, while $c$ is the sound velocity and $k_F$ is the Fermi quasi-wave vector.

After the diagonalizations of Exp.(3.5), using Bogoliubov transformations[12] the following relation between the order parameters at zero–temperature and non zero-temperature is obtained

$$\ln\frac{\Delta(0)}{\Delta(T)} = 2\sum_{n=1}^{\infty} (-1)^{n+1} K_0\left(n\frac{\Delta(T)}{\kappa_B T}\right) - \frac{1}{\lambda}\frac{\gamma}{1+\gamma} , \quad (3.7)$$

where: $K_0$ is Mackdonald's function; $\lambda = |G^{(0)}| v_F$; $\Delta(0) = 2\hbar\omega_{max} \exp(-1/\lambda)$ and $v_F = m^* k_F /(2\pi^2 \hbar^2)$ is the density of states on the Fermi surface. In the limit cases the Exp. (3.6) takes the following values:

$$\gamma = \begin{cases} \gamma_1, for \chi \gg 1 \\ \gamma_2, for \chi \ll 1 \end{cases}; \quad \gamma_1 = \sigma T^2, \sigma = \frac{\pi^2}{3}\left(\frac{2\kappa_B}{\hbar c k_F}\right)^2 \left(\ln\left|\frac{q_{max}}{k_F}\right|\right)^{-1}; \quad \gamma_2 = 2\left(\ln\left|\frac{q_{max}}{k_F}\right|\right)^{-1} \quad (3.8)$$

The numerical solution of Eq.(3.7) is obtained using the following values of the parameters $k_F \approx 2\cdot 10^7 cm^{-1}, q_{max} \approx 1\cdot 10^8 cm^{-1}, c = 1\cdot 10^5 cm/s, \rho = 3g/cm^3, \varepsilon_{21} \approx 3 eV, m^* \approx 5m_e$ ($m_e$ is the mass of free electron). The comparison of dependence of order parameter as function of temperature $\Delta(T)$ with BCS-like model with same $\Delta(0)$ is given in Fig.1. The critical temperature can be found from the relation

$$\ln\left(\frac{\pi\kappa_B T_c}{\Delta(0)}\right) = \frac{1}{\lambda}\frac{\sigma T_c^2}{1+\sigma T_c^2} + C_0, \quad C_0 = 0.577... \text{ - Euler constant.} \quad (3.9)$$

In order to estimate the power law dependence of order parameter near the critical temperature $T_c$ one can consider the Mackdonald's function to integral form and decompose this integral on the small parameter $\Delta(T)/(\kappa_B T)$. Taking into account the Eq.(3.9) for $T_c$ it is obtain the following temperature dependence of the order parameter

$$\Delta(T) = \frac{4\pi\kappa_B T}{\sqrt{14\xi(3)}}\left[\ln\frac{T}{T_c} + \frac{1}{\lambda}\left(\frac{\sigma T^2}{1+\sigma T^2} - \frac{\sigma T_c^2}{1+\sigma T_c^2}\right)\right]^{1/2}, \quad (3.10)$$

where $\xi(3)$ is the value of Riemann Zeta-function ($\xi(3) \approx 1.2$).

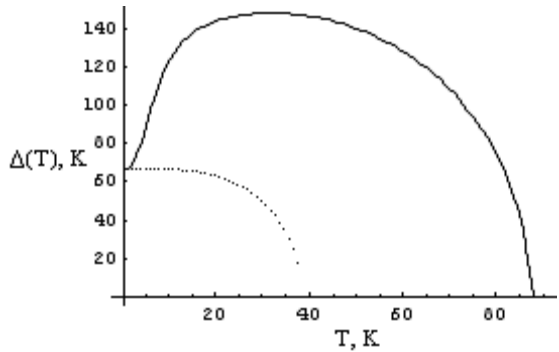
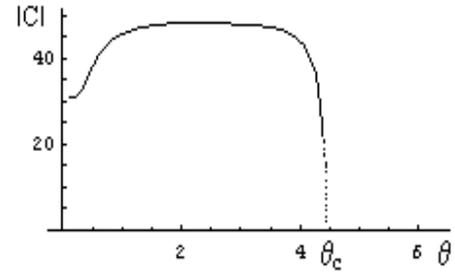

Fig.1 The temperature dependence of superconductivity order parameter $\Delta(T)$ (······BCS-like theory; ——Two-phonon superconductivity theory)

Fig.2 The temperature dependence of two-photon super-radiance order parameter $|C|$ (For $N=100$; $|\chi/\omega_0| = 2.4/N$)

## 4. Results and Discussions

In this paper the analogy between the two-photon cooperative correlation of the atoms in micro-cavities and the formation of Cooper-pairs in the processes of the two-phonon electron-electrons interaction was discussed. It is shown that the exchange integral between atoms or electrons increases with temperature due to the fact that the two-quantum scattering effect stimulates the coherence formation of atomic polarization and Cooper-pairs in the superconductor. This effect opens a new concept on the formation of new correlation phase and new behavior of phase transitions in the processes of multi-quantum exchange between the atoms or carriers. It is clear that with the increase of the temperature, the atomic polarization and the superconductivity gap increases slower than $k_B T$ so that the order parameter at low temperature achieves the maximum value and after decreases till zero in critical temperature. This anomalous dependence on the temperature of the order parameter is described by a complicated two-quantum exchange between the atoms or electrons.